\documentstyle[prl,aps,preprint,epsf]{revtex}

\begin{document}

\widetext
\draft

\title {An Exact Monte Carlo Method for Continuum Fermion Systems}

\author{M. H. Kalos$^1$ and Francesco Pederiva$^2$} 
\address{$^1$Lawrence Livermore National Laboratory
 Livermore, CA 94550 USA}
\address{$^2$Dipartimento di Fisica, Universit\`a di Trento,
I-38050 Povo-Trento, Italy
} 
\date{\today}
\maketitle

\begin{abstract}

We offer a new proposal for the Monte Carlo treatment of many-fermion systems in
continuous space.  It is based upon Diffusion Monte Carlo with significant
modifications:  correlated pairs of random walkers that carry opposite signs;
different functions ``guide'' walkers of different signs; the Gaussians used for
members of a pair are correlated; walkers can cancel so as to conserve their
expected future contributions.  We report results for free-fermion systems and a
fermion fluid with 14 $^3$He atoms, where it proves stable and correct.  Its
computational complexity grows with particle number, but slowly enough to make
interesting physics within reach of contemporary computers.

\end{abstract}

%\begin{multicols}{2}

%\section{Introduction} 

Monte Carlo methods have provided powerful numerical tools for quantum many-body
physics \cite{CepKal79,SchKal84}.  They include methods such as Green's function
Monte Carlo (GFMC) \cite{KLV}, Diffusion Monte Carlo (DMC)\cite{RCAL}, or Path
Integral Monte Carlo (PIMC)\cite{PIMC} that are capable of giving, at least for
moderate size bosonic systems, answers with no uncontrolled approximations.
Such accurate treatment of fermionic systems has been made vastly more difficult
by a ``sign problem.'' Progress in the application of Quantum Monte Carlo
methods to condensed matter physics, to electonic structure, and to nuclear
structure physics has been impeded for years by the lack of exact and efficient
methods for dealing with fermions.

This paper offers a new proposal for solving many-fermion systems by an
extension of DMC.  In the systems we have studied, the signal-to-noise ratio of
the Monte Carlo estimates are constant at long imaginary times, by contrast to
the behavior of ordinary DMC where they decay exponentially \cite{SchKal84}.
Except for the use of a short-time Green's function, no approximations--
physical, mathematical, or numerical-- are made.  The effect of a finite
interval of imaginary time is easily controlled, or may be eliminated entirely.

It is no surprise that Monte Carlo methods can solve the many-body Schrodinger
equation in imaginary time for bosonic systems.  Let $\vec R$ denote all
coordinates of an $N$-body system, and $V(\vec R)$ be the potential at $\vec R$.
\begin{equation} \label{eqI.0} [-\frac{\hbar^2}{2m} \nabla^2 +V(\vec R)]
\psi(\vec R,\tau) \\ + \hbar \frac{\partial \psi(\vec R,\tau)}{ \partial \tau} =
0 \end{equation}

This equation also describes the diffusion of an object (a ``random walker'') in
a $3N$ dimension space in which the potential $V(\vec R)$ serves as a
generalized absorption rate.  Because the potential in physical problems can be
unbounded from above and below, a direct simulation of that diffusion, although
straightforward, will be inefficient.  Some form of importance sampling
transformation has been found to be highly useful.  In the standard
DMC\cite{KLV,RCAL}, this is a technical device for accelerating the convergence;
in our new method it becomes an essential feature.

DMC uses an ``importance'' or ``guiding'' function $\psi_G(\vec R)$ and a trial
eigenvalue $E_T$ to construct a random walk.  A simple version is as follows:
Using a fixed step in imaginary time, $\delta \tau$, a walker at $\vec R$ is (a)
moved to $\vec R + \delta \tau \vec \nabla \ln \psi_G(\vec R)$; (b) then each
coordinate is incremented by an element of a random vector $\vec U$, a Gaussian
with mean zero and variance $\delta \tau$; finally, (c) each walker is turned
into $M$ walkers with $<M>=\exp\left \{ \delta \tau \left [E_T - {\hat H
\psi_G(\vec R)}/{\psi_G(\vec R)} \right ] \right \} $, where $\hat H$ is the
Hamiltonian.

The resulting random walk has expected density \begin{equation} \label{eqI.1}
f(\vec R,\tau)=\psi_G(\vec R) \sum_k a_k \exp [(E_T - E_k) \tau ] \phi_k(\vec R)
\end{equation} where $\phi_k(\vec R)$ are eigenfunctions of $\hat H$ with
eigenvalues $E_k$, and $a_k$ are expansion coefficients.  The limit of $f(\vec
R,\tau)$ for large $\tau$ is dominated by the eigenfunction $\phi_0$ having the
lowest eigenvalue $E_0$.

We alter the structure of DMC in the following ways:  (1) In order to represent
an antisymmetric wave function that is both positive and negative, we introduce
walkers, $\{\vec R^+_m,\vec R^-_m \}$, that respectively add or subtract their
contributions to statistical expectations.  The computation now involves
ensembles of pairs of walkers carrying opposite signs.  (2)Two distinct
functions, $\psi_G^{\pm}(\vec R)$ are used to guide the $\pm$ walkers.  (3) The
Gaussians $\vec U^{\pm}$ for the paired walkers are not independent; rather
$\vec U^-$ is obtained by reflecting $\vec U^+$ in the perpendicular bisector of
the vector $\vec R^+ -\vec R^-$.  Finally (4) the overlapping distributions that
determine the next values of $\vec R^{\pm}$ are added algebraically so as to
allow positive and negative walkers to cancel, while preserving the correct
expected values.

In a Monte Carlo calculation of this kind, we ``project'' quantities of interest
by calculating integrals weighted with some trial function, say $\psi_T(\vec
R)$.  In DMC the energy eigenvalue, $E_0$, can be determined from:

\begin{equation}
\label{eqI.5}
E_0 = \frac {\int \hat H \psi_T(\vec R) \phi_0(\vec R) d\vec R}
{\int \psi_T(\vec R) \phi_0(\vec R) d\vec R} =
\frac 
{\displaystyle\sum_m \frac {\hat H \psi_T(\vec R_m)}{\psi_G(\vec R_m)} }
 {\displaystyle \sum_m 
\frac {\psi_T(\vec R_m)}{\psi_G(\vec R_m)} }
\end{equation}
replacing integrals by sums over positions of the random walk.

In our modified dynamics, Eq.(\ref{eqI.5}) now takes the form
\begin{equation}
\label{eqI.6}
E_0 = \frac {\displaystyle \sum_m [\frac {\hat H \psi_T(\vec R^+_m)}
{\psi_G^+(\vec R^+_m)} 
- \frac {\hat H \psi_T(\vec R^-_m)}{\psi_G^-(\vec R^-_m)}] }
 {\displaystyle\sum_m [ \frac {\psi_T(\vec R^+_m)}{\psi_G^+(\vec R^+_m)}
 - \frac {\psi_T(\vec R^-_m)}{\psi_G^-(\vec R^-_m)}] }.
\end{equation}

%\section{Correlated Pairs}

If $\{\vec R^+_m \}$ and $\{ \vec R^-_m\}$ follow the same dynamics using the
same $\psi_G$, then the expected values of numerator and denominator of Eq.
(\ref{eqI.6}) decay exponentially at large $\tau$.  Some correlation among
walkers is essential.  This observation is reinforced by noting that the Pauli
principle, which demands that fermion wave functions be antisymmetric, is a
global condition, and cannot be satisfied by independent walkers.  Put another
way, it will be necessary to have dynamics that distinguish between walkers that
carry different signs.  These motivations underlie aspects (2) and (3) of our
method outlined above.

A second aspect of the difficulty in treating fermion systems is that the
density that one obtains naturally from a random walk is the symmetric ground
state.  In order for Eq.  (\ref{eqI.6}) to have an asymptotically bounded
signal-to-noise ratio, walkers of opposite signs must be able efficiently to
cancel.  This underlies the need for modification (4) given above.  The need for
some degree of cancellation has been a theme of previous research starting with
the work of Arnow et al.  \cite{Arnow}.  The need for distinct dynamics for
positive and negative walkers was stressed in \cite{ZK91}.  That these two aims
could be accomplished by appropriate correlation among walkers was pointed out
by Liu, Zhang, and Kalos \cite{LZK}.  The use of distinct guiding functions is
new and serves as the connection among the different algorithmic ideas that
enables the treatment of general potentials.

Stable results can be obtained using correlated pairs only.  Correct results are
ensured when the dynamics have the property that the random walk for either
member of the pair is the same as that of a single free walker, except when they
cancel, a condition satisfied here.  The expectations of Eqs.(\ref{eqI.5}) and
(\ref{eqI.6}) are linear in the walker densities and are unchanged by
correlations.  Furthermore, we have devised a method of canceling opposite
walkers that also preserves these expectations.

%\section{The Stochastic Dynamics}
Let $\varphi_A(\vec R)$ be a trial function for the fermionic state.  
Let $\varphi_S(\vec R)$ be some approximation to the symmetric ground state
wave function of the same Hamiltonian.
Define:
\begin{equation}
\psi_G^{\pm}(\vec R) = \sqrt {\varphi_S^2(\vec R) + c^2 \varphi_A^2(\vec R)
} \pm c\varphi_A(\vec R)
\end{equation}

The following properties of these two functions are significant:  (a) they are
positive; (b) when $c$ is small, they are dominated by $\varphi_S$, so that
opposite walkers behave similarly; (c) $\psi_G^+$ transforms under an odd
permutation $\cal P$ as follows:

\begin{equation}
\label{eqa.0}
\psi_G^+({\cal P} \vec R) = \psi_G^-(\vec R) .
\end{equation}
As mentioned above we modify simple DMC in several ways.  The ``drift''
is applied in the usual way to walkers assumed to be at $\vec R^{\pm}_0$,
using the two guiding functions:
\begin{eqnarray}
\label{eqa.1}
\begin{array}{c}
\vec R^+ = \vec R^+_0 + \delta \tau \vec \nabla \ln \psi_G^+(\vec R^+) \\ \\
\vec R^- = \vec R^-_0 + \delta \tau \vec \nabla \ln \psi_G^-(\vec R^-) .
\end{array}
\end{eqnarray}
Diffusion  of the  walkers,  however, is carried out in a  correlated  way:  let
$\vec U^+$ be a vector of $3N$ Gaussian  random  variables each of mean zero and
variance $\delta \tau$.  New trial positions $\vec{R}_n^\pm$ are now given by
\begin{equation}
\label{eqa.7}
\vec R^+_n = \vec R^+ + \vec U^+ ; \vec R^-_n = \vec R^- + \vec U^- ,
\end{equation}
where the random vector $U^-$ is obtained by reflection in the perpendicular
bisector of the vector $\vec R^+ -\vec R^-$ as described in (4) above.

Walker densities can be subtracted by computing their chances of
arrival at a common point, but because they have different guiding functions,
they do not exactly cancel.  The analysis of ``forward walking''
\cite{CepKal79,Kal70,FW91} allows one to determine the expected future
contribution of a walker to any projected quantity.  Thus, we can compute the
change in expectations when a pair meets.  For this change to be zero
\cite{nato}, a positive walker at $\vec R^+_n$ must survive to the next time
step with probability

\begin{eqnarray}
\label{eqa.11}
\begin{array}{c}
P^+(\vec R^+_n ; \vec R^+,\vec R^-) = \\ \\
  \max \left [0,1- \frac{\displaystyle B^-(\vec R^+_n | \vec R^-) 
  G(\vec R^+_n-\vec R^-) \psi^+_G(\vec R^+_n)}
{\displaystyle
B^+(\vec R^+_n | \vec R^+) G(\vec R^+_n- \vec R^+) \psi^-_G(\vec R^+_n)}  
\right ]
\end{array}
\end{eqnarray}
where
\begin{equation}
\label{eqa.8}
G(\vec R^{\prime}-\vec R)=
\frac{\exp[-(\vec R^{\prime}-\vec R)^2/(2 \delta \tau)]}
{(2 \pi \delta \tau)^{3N/2}}
\end{equation}
is the Gaussian density used in DMC.  The branching factors, $B^+(\vec R | \vec
R^+)$ and $B^-(\vec R | \vec R^-) $ are
\begin{equation}
\label{eqa.13}
B^{\pm }(\vec R) = \exp\left \{ \delta \tau \left [E_T - 
\frac {H \psi^{\pm }_G(\vec R)}{\psi^{\pm }_G(\vec R)} \right ] \right \}
\end{equation}
An analogous expression is used for negative walkers.

An isolated walker may appear as a result of different branching factors at
$\{\vec R^+_m \}$ and $\{\vec R^-_m \}$; if, with probability one half, one
generates a walker of opposite sign by interchanging the coordinates of two
like-spin particles, then a pair is reconstituted that preserves future
expectations.

To determine the energy, we use the estimator of Eq.  (\ref{eqI.6}).  A sharp
indication of the stability of the calculation is the behavior of its 
denominator
\begin{equation}
{\cal D} = [ \frac {\psi_T(\vec R^+_m)}{\psi_G^+(\vec R^+_m)}
 - \frac {\psi_T(\vec R^-_m)}{\psi_G^-(\vec R^-_m)}] 
\end{equation} 
In a naive calculation, $\cal D$ decays to zero in an
imaginary time of order $\tau_c=1/(E_A-E_S)$ where $E_A$ and $E_S$ are the
fermion and boson energies. A stable method will show $\cal
D$ asymptotically constant.

Although a system of free fermions in a periodic box is analytically trivial, it
presents an exigent test of this method.  For this system, the lowest symmetric
state is constant, and the exact fermionic wave function is a determinant of
plane waves.  We use $\rho=0.5$ and set

\begin{equation}
\psi_G^{\pm}(\vec{R})=\sqrt{1+c^2\varphi_A^2(\vec{R})
}\pm c\varphi_A(\vec{R})
\end{equation}
where $\varphi_A$ is a Slater determinant of one body orbitals
$\chi^{\vec{k}}_{\vec{r}_i}$ of the following form:
\begin{equation}
\chi^{\vec{k}}_{\vec{r}_i}=\exp\left[i\vec{k}\cdot\left(\vec{r}_i
+\lambda_B\sum_{j\ne i}\eta(r_{ij})\vec{r}_{ij}\right)\right]
\end{equation}
The parameter $\lambda_B$ controls the departure of the nodal structure of
this function from the exact shape.  The fact that these functions are modulated 
only a little from a constant by $\varphi_A$ means that the polarization of the
population of plus and minus walkers is small.

In table I we report the results obtained for periodic systems of 7, 19, and 27
free fermions.  The results agree with the analytic eigenvalues within the Monte
Carlo estimates of the standard error.  It has been conjectured that the
computational complexity of Fermion Monte Carlo calculations will grow as $N!$,
where $N$ is the number of particles in the system.  Since (27!/7!)  = 2.16
$\times 10^{24}$, a calculation with 27 or even 19 bodies would be impossible
were that conjecture to be true.

We have also applied this algorithm to a system of 14 $^3$He atoms in a periodic
box at equilibrium density, $\rho=0.0216$\AA$^{-3}$.  Energies are expressed in
Kelvins, and lengths in \AA.

With interatomic potentials that have a hard core, we may use the same
function $\varphi_A$ as for free fermions, but also need a Jastrow
product.  With 
\begin{equation}
\varphi_S = \varphi_S(\vec{R})=\prod_{i<j}\exp[-(b/r_{ij})^5],
\end{equation}
the guiding functions now have the form:
\begin{equation}
\psi_G^{\pm}(\vec{R})=\varphi_S(\vec{R})\left[\sqrt{1+c^2\varphi_A^2(\vec{R}) }
\pm c  \varphi_A(\vec{R})\right]
\end{equation}

In Fig. 1 we plot the cumulative denominator as a function of imaginary time
for a typical run.  As can be seen, the fundamental stability of the method is
well demonstrated.  Fig.  2 shows the decay of the same denominator, when the
method is made unstable by setting $c=0$.  

Table II exhibits the eigenvalues of various runs with our method applied to the
periodic system with 14 $^3$He atoms.  They are all consistent, and yield a
weighted average of -2.2558(39).  The run marked (b) is a continuation of the
run labeled (a) separated by a long run with a longer time step.  As a whole,
including such continuations, the longest aggregate sequence comprises a total
imaginary time of 1830 $K^{-1}$.  Using a total system energy difference of 20
$K$ (as we have measured), that corresponds to $3.6\times 10^4$ fermion decay
times.  An alternative measure of the length of the run, suggested by David
Ceperley\cite{CepPC}, is the ratio of the rms diffusion length of a particle to
the mean spacing between particles.  For this sequence of runs, that ratio is
19.  Thus the observation of stable values of the sums in Eq. (\ref{eqI.6}) is
significant.

Space limitations preclude a complete description of the other checks that we
have made that the results for $^3$He are correct:  they include a fixed node
calculation of exactly the same model problem, which yielded an eigenvalue of
-2.08(1) $K$.  A transient estimate (cf.  Fig.  3), relaxing from the fixed
node, is consistent with our result (shown as the dashed line.)  Analysis of the
results in Fig. 2 leads to a fermion-boson energy difference of 1.434(35) $K$
per particle.  This agrees well with a direct calculation of the energy of a
14-body mass-3 boson system that gave -3.68(1) $K$.

%\section{Conclusions} 

By construction, the method proposed here introduces no approximations other
than that of the short imaginary-time Green's function.  In other words, if the
results are stable, then they are correct.  Although we have have not yet proved
the stability of the method (i.e.  that the long-term average of the denominator
of the eigenvalue quotient is not zero), we believe that we have convincingly
demonstrated the stability.  Perhaps the most important conclusion that we may
draw is that the ``sign problem'' of Fermion Monte Carlo for continuous systems
is not intractable; the search for elegant computational methods in this and
related applications is justified.

%\section*{Acknowledgments}

One of the authors (FP) has been supported, in part, by the National Science
Foundation under grant ASC 9626329.  This work was performed under the auspices
of the U.S. Department of Energy by Lawrence Livermore National Laboratory
under contract No.  W-7405-Eng.-48.  We are happy to acknowledge helpful
conversations with B.J. Alder, J.  Carlson,  D.M. Ceperley,  G.V. Chester,
R.Q. Hood, S.B. Libby, K.E. Schmidt, C.J. Umrigar, and S. Zhang.  We thank
particularly D.E. Post for his help and strong support.

%\end{multicols}

%\begin{multicols} {2}
\begin{table}
\begin{tabular}{ccc}
   N&    $E$ &   $E_{ex}$\\
\hline
   7&   2.91285(49)&     2.912712 \\
  19&   2.760(25) &      2.757454  \\
  27&   2.796(30) &      2.763316
\end{tabular}
\caption{Energies and errors for a periodic system
of $N$ free fermions. The analytic result is $E_{ex}$.}
\end{table}
%\end{multicols}
%
% TABLES
%
\begin{table}
\begin{tabular}{cccc}
	  b&	  c&   	$\lambda_B$&	    $E$	\\
\hline
   	1.15&	0.025&	 0&	-2.251(08)a	\\
   	1.145&	0.025&	 0&	-2.258(17)  	\\
   	1.135&	0.025&   0&	-2.257(10) 	\\
   	1.15&	0.016&	 0&	-2.246(20)  	\\
   	1.15&	0.010&	 0&	-2.250(19)  	\\
	1.15&	0.025&   0&	-2.2559(84)b	\\
  	1.15&	0.025&  -0.05&	-2.249(12)      \\
  	1.15&	0.025&   0.05&	-2.268(10)	  
\end{tabular}
\caption{Energies and errors for a periodic system
of 14 $^3$He atoms}
\end{table}
%
% FIGURES
%
\begin{figure}
%\vspace{1.3cm}\noindent
%\centerline{\epsfxsize=15.0cm  \epsfbox{Mel1po.eps}}
\caption{Cumulative denominator of energy quotient: a stable calculation}
\end{figure}

\begin{figure}
%\vspace{1.3cm}\noindent
%\centerline{\epsfxsize=15.0 cm \epsfbox{Mel2po.eps}}
\caption{Denominator of energy quotient: an unstable calculation.
The smooth curve is $\exp(-20.08 \tau)$, a fit to the data.}
\end{figure}

\begin{figure}
%\vspace{1.3cm}\noindent
%\centerline{\epsfxsize=15.0cm  \epsfbox{te.eps}}
\caption{Relaxation of eigenvalue from fixed node: a transient calculation.}
\end{figure}

%\end{multicols}
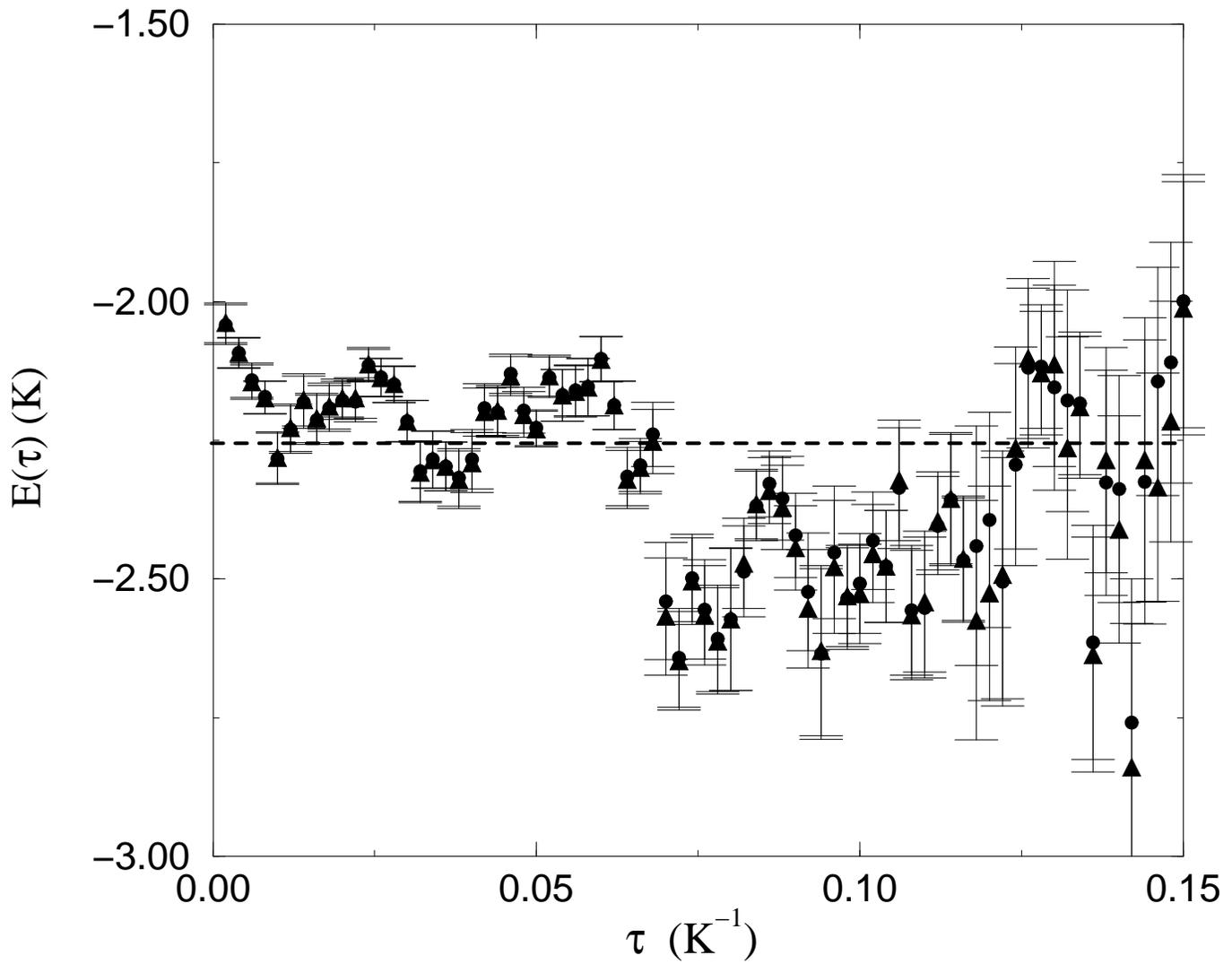
\end{document}